\documentclass[10pt]{article}

\usepackage{amsmath}
\usepackage{array}
\usepackage{appendix}
\usepackage{graphicx}
\usepackage{amsfonts}
\usepackage{amssymb}
\usepackage{mathrsfs}
\usepackage{yfonts}
\usepackage{euscript}
\usepackage{upgreek}
\usepackage{slantsc}
\usepackage{calligra}
\usepackage[T1]{fontenc}
\usepackage{epsf}
\usepackage{latexsym}

\usepackage{tipa}

\def\be{\begin{equation}}
\def\ee{\end{equation}}
\def\beq{\begin{equation}}
\def\eeq{\end{equation}}
\def\bea{\begin{eqnarray}}
\def\eea{\end{eqnarray}}

\def\ni{\noindent}

\def\!{\hspace{-1.6667em}}

\def\mD{\mbox{D}}

\def\mG{\mbox{G}}
\def\mH{\mbox{H}}

\def\mK{\mbox{K}}

\def\mP{\mbox{P}}
\def\mQ{\mbox{Q}}
\def\mR{\mbox{R}}
\def\mS{\mbox{S}}
\def\mT{\mbox{T}}

\def\me{\mbox{e}}

\def\mh{\mbox{h}}

\def\mn{\mbox{n}}   
\def\mo{\mbox{o}}
\def\mp{\mbox{p}}

\def\mt{\mbox{t}}

\def\bupSigma{\mbox{\boldmath$\Sigma$}}                 

\def\se{\mbox{\scriptsize e}}

\def\sh{\mbox{\scriptsize h}}

\def\sll{\mbox{\scriptsize l}}  
\def\sm{\mbox{\scriptsize m}}
\def\sn{\mbox{\scriptsize n}} 
\def\so{\mbox{\scriptsize o}} 
\def\sp{\mbox{\scriptsize p}}

\def\sA{\mbox{\scriptsize A}} 
\def\sB{\mbox{\scriptsize B}}

\def\sD{\mbox{\scriptsize D}}

\def\sM{\mbox{\scriptsize M}}

\def\sS{\mbox{\scriptsize S}}
\def\sT{\mbox{\scriptsize T}}

\def\sV{\mbox{\scriptsize V}}
\def\sW{\mbox{\scriptsize W}}

\def\te{\mbox{\tiny e}}

\def\ti{\mbox{\tiny i}}

\def\tn{\mbox{\tiny n}}

\def\tr{\mbox{\tiny r}}

\def\ttt{\mbox{\tiny t}}   
\def\tu{\mbox{\tiny u}}

\def\tA{\mbox{\tiny A}}

\def\tD{\mbox{\tiny D}}

\def\tM{\mbox{\tiny M}}

\def\tV{\mbox{\tiny V}}

\def\K{Kucha\v{r} }

\def\pa{\partial}
\def\d{\textrm{d}}

\def\5Star{\mbox{\Large$\star$}}              
\def\sumi2{\sum\mbox{}_{\mbox{}_{\mbox{\scriptsize $i$=1}}}^2}
\def\sumi3{\sum\mbox{}_{\mbox{}_{\mbox{\scriptsize $i$=1}}}^3}

\def\sumnlm{\sum\mbox{}_{\mbox{}_{\mbox{\scriptsize $\sn, \sll, \sm$}}}}

\def\sumj3{\sum\mbox{}_{\mbox{}_{\mbox{\scriptsize $j$=1}}}^3}
\def\sumk3{\sum\mbox{}_{\mbox{}_{\mbox{\scriptsize $k$=1}}}^3}

\begin{document}

\begin{center}

\Large{\bf Origin of Structure in the Universe: Quantum Cosmology Reconsidered}\normalsize

\vspace{.15in}

{\large \bf Edward Anderson} 

\vspace{.15in}

\large {\em DAMTP, Centre for Mathematical Sciences, Wilberforce Road, Cambridge CB3 OWA.  } \normalsize

\end{center}

\begin{abstract}
 
Based on a more careful canonical analysis, we motivate a reduced quantization -- in the sense of superspace quantization -- 
of slightly inhomogeneous cosmology in place of the Dirac quantization in the existing literature, and provide it in the vacuum case.
This is attained through consideration of configuration space geometries at various levels of reduction.
Some of these have the good fortunate of being flat.
Geometrically natural coordinates for these are interpreted in terms of the original redundant formulation's well-known mode expansion coefficients.  
 
\end{abstract}

\section{Introduction}

I consider perturbatively inhomogeneous quantum cosmology models about the spatially closed $\mathbb{S}^3$ isotropic model, 
similar to the well-known one of Halliwell and Hawking \cite{HH85}. 
Such models are important as a candidate for the origin of structure in the universe: quantum cosmological fluctuations as amplified by inflation.  
The type of semiclassical regimes in \cite{HH85} represent a more general setting for this than in purely QFT approaches, 
providing both foundational justification for the latter and more detailed calculations. 
This slightly inhomogeneous cosmology is additionally an interesting model arena \cite{AHH-1} for Background Independence and the Problem of Time in Quantum Gravity 
\cite{Kuchar92I93, APoT2, APoT3}.  
Whereas the Frozen Formalism Problem often associated with the Wheeler--DeWitt equation is the most widely known Problem of Time facet, there are many others 
and they are heavily interconnected. 
Thus model arenas exhibiting between most and all of these features -- while managing to be substantially more tractable than full GR -- are of interest.

\mbox{ } 

\ni Furthermore, a number of unusual features arose in this study in the case including minimally-coupled scalar field matter \cite{AHH-1}.    
Some of these features apply also to \cite{HH85}'s model itself.  

\mbox{ }

\ni Observation 1) is that upon applying the SVT (scalar--vector--tensor) split to the GR Hamiltonian constraint ${\cal H}$ and the GR momentum constraint ${\cal M}_i$.  
Whereas ${\cal H}$ and ${\cal M}_i$ are a priori first class constraints, the SVT split pieces of these turn out not to be first-class constraints.

\mbox{ } 

\ni Consequently, proceeding to quantization by just hatting these split pieces as if they were first-class constraints becomes a questionable step to take.  
Thus observation 1) invites a re-analysis of i)  the quantization step of \cite{HH85} (as well as of a first analysis of \cite{AHH-1}'s quantization), 
                                         and ii) of subsequent semiclassical work.  
[ii) is still expected to involve the usual harmonic oscillator type mathematics {\sl at lowest nontrivial order}.] 

\mbox{ }

\ni Also, a particular view was taken in \cite{AHH-1} as regards which quantities to solve ${\cal M}_i$ for.
I.e. the {\it thin sandwich} prescription \cite{WheelerGRT, San-Concat, TSC2}, in which the Lagrangian variables form of ${\cal M}_i$ is to be solved for the shift vector $\upbeta_i$.  
This choice was made due to the role of the Thin Sandwich Problem as a second facet of the Problem of Time in Quantum Gravity.

\mbox{ }

\ni Observation 2) is that in perturbatively inhomogeneous cosmology to second order, 
one has the good fortune that the thin sandwich elimination -- a type of reduction -- involves just very simple algebra.
Moreover, the resultant `reduced configuration space' geometry ensuing for the full Einstein--scalar field system case of \cite{AHH-1} turns out to be of the wrong dimension. 
[It is 8-$d$ rather than 6-$d$ for each mode's separated out problem, i.e. a discrepancy of 2.]
This indicates that the procedure has ceased to fully eliminate the spatial diffeomorphisms, Diff($\bupSigma$).  
This obstructs progress with careful assessment of how to quantize the full Einstein--scalar field case.

\mbox{ }

\ni New Observations 3) and 4) are that the above two observations also have vacuum counterparts.
Furthermore, now they have the notable distinction of being simpler 
-- simple enough to resolve them, and as such I wrote this new Paper covering this simpler case for which those workings could be completed.
This requires some discussion of the Dirac Algorithm in the Principles of Mechanics, so I outline this in Section 2.
In Section 3, I introduce the Baierlein--Sharp--Wheeler (BSW) \cite{BSW} action. 
I use this due to its implementing Background Independence and Problem of Time resolving properties 
(e.g. as a classical precursor of the quantum-level Frozen Formalism Problem).  
Since I use the BSW action, it is useful for me to explain how the GR Hamiltonian constraint ${\cal H}$ can change status from secondary to primary, 
so that these are in fact formulation-dependent adjectives rather than just absolutes within a given theory.
I support this with Appendix A on keeping track of degrees of freedom counts.
Using the BSW action also requires assessing some caveats (Appendix B).
I then explain in Section 4 how slightly inhomogeneous cosmology already exhibits all the other Background Independence aspects and Problem of Time facets that GR does, 
with the benefit of many of these being perturbatively small and hence more tractable.
This completes the incipient motivation for this study.
Details of the model's SVT split are presented in Appendix C. 
In Section 5 then consider how first-class and second-class are also formulation-dependent adjectives as applied to constraints, and can furthermore be influenced by applying manoeuvres 
such as the SVT split.
This then enters my explanation of how not all the SVT split pieces of the vacuum case's incipient first-class constraints are themselves first-class constraints.

\mbox{ } 

\ni Section 6 outlines the Thin Sandwich Problem, which itself is a second PoT facet.
Section 7 shows how this greatly simplifies for slightly inhomogeneous cosmology, 
though it does not by itself attain the amount of reduction expected (to superspace).
The current Paper's vacuum case of thin sandwich reduction produces a 5-$d$ reduced configuration space rather than the expected 4-$d$ one, i.e. now a discrepancy of just 1.  
I can furthermore show how to complete that reduction in this case. 
This involves reducing out the V-piece of ${\cal H}$ as well 
(with reference to Appendix D's account of the new variables which arise during the reduction procedure) and thus one indeed arrives at a 4-$d$ configuration space.
Thus for this model reduced quantization (in the sense of removing linear constraints) and superspace quantization \cite{I76} cease to coincide (Section 8).  
In Section 9, I end by setting up superspace quantization for this model, by making use of the previous Sec's reduced geometry.

\section{Overview of canonical theory of constrained systems}

One way in which systems of fundamental physics equations can be nontrivial is through their possession of constraints.
Constraints as manifested within this setting are usefully characterized in canonical terms.
Here the basic variables are configurations and momenta, and Poisson brackets structure is available.
Then the constraints in question are relations between the fundamental physics theory in question's momenta, by which these are not all independent.
The canonical approach further affords some useful further characterizations of constraints -- and other entities arising alongside them -- as follows \cite{Dirac, HTbook}.  

\mbox{ } 

\noindent {\it Secondary constraints} arise through use of the equations of motion.    

\noindent {\it Primary constraints}, on the other hand, are those which arise purely from the form of the Lagrangian; see Section 3 for examples.

\noindent Dirac's notion of {\it weak equality} is equality up to additive functionals of the constraints.  
This holds on the so-called {\it constraint surface}: the surface within phase space where the all the constraints vanish.

\mbox{ }  

\ni A distinct classification of constraints is into {\it first-class} constraints: those whose classical brackets with all the other constraints vanish weakly.  
[At least at the outset, these classical brackets are the above-mentioned Poisson brackets.]
 
\ni {\it Second-class} constraints are then simply defined to be those which are not first-class.  
Note for degrees of freedom counting purposes that whereas first-class constraints use up two degrees of freedom each, second-class ones use up only one \cite{HTbook}.

\mbox{ } 

\ni Dirac begins to handle constraints by additively appending them with Lagrange multipliers to the incipient Hamiltonian for the system.
The {\it Dirac Algorithm} \cite{Dirac} involves checking whether a given set of constraints implies any more constraints or any further types of entity.
The entities arising thus can be of five types.

\mbox{ } 

\ni Entity 0) {\it Inconsistencies}.

\ni Entity 1) Mere {\it identities} -- equations that reduce to $0 \approx 0$, i.e. $0 = 0$ modulo constraints.

\ni Entity 2) Equations independent of the Lagrange-multiplier unknowns, which constitute extra secondary constraints.

\ni Entity 3) Demonstration that some of the constraints known so far are in fact second-class, 
by being second-class with respect to subsequently found constraints implied by the original constraints.

\ni Entity 4) Relations amongst some of the appending Lagrange multipliers functions themselves, which are a further `{\it specifier equation}' type of equation 
(i.e. specifying restrictions on the Lagrange multipliers).\footnote{Dirac introduces these entities on p 14 of \cite{Dirac}; he uses `imposes a condition' 
in pointing out the existence of specifier equations; the term `fixing equations', as in e.g. lapse-fixing equation, is often used for such in Numerical Relativity.}

\mbox{ }

\ni Lest entity 0) be unexpected, Dirac supplied a basic counterexample to Principles of Dynamics formulations entailing consistent theories.  
For Lagrangian $L = x$, the Euler--Lagrange equations read $0 = 1$. 
If entity 0) occurs, the candidate theory is wrong. 
Thus one either simply gives up on it or one modifies the incipient Lagrangian to pass to a further theory for which this does not happen.
Then call entities 1) to 4) the `consistent entities' arising from the Dirac Algorithm.
Moreover, since entity 1) is an equation with no new content, call entities 2) to 4) the `nontrivial consistent entities.  
This is a generalization from considering the set of constraints to considering the set of constraints {\sl and specifier equations arising alongside them} in the Dirac Algorithm.  

\mbox{ } 

\ni The Dirac Algorithm is to be applied recursively until one of the following three conditions holds.

\mbox{ }

\ni Termination 0) {\it Inconsistent theory}, due to a case of 0) arising.

\ni Termination 1) {\it Trivial theory}, due to the iterations of the Dirac Algorithm having left the system with no degrees of freedom. 

\ni Termination 2) {\it Completion}. Here the latest iteration of the Dirac Algorithm has produced no new nontrival consistent entities, indicating that all of these have been found. 

\mbox{ } 

\ni Ways of handling nontrivial consistent entities other than first-class constraints are also pertinent.
This is especially so since many standard quantum procedures are based on just first-class constraints remaining by that stage.
This usually entails classical removal of any other nontrivial consistent entities which are present in the original formulation.
Four different approaches to this are as follows.

\mbox{ }

\ni Procedure A) Replace the incipient Poisson brackets with Dirac brackets \cite{Dirac}, by which second-class constraints are removed.

\ni Procedure B) Extend the phase space with further auxiliary variables so as to `gauge-unfix' second-class constraints into first-class ones. \cite{BT91, HTbook}. 

\ni Procedure C) Classically reduce out the entities in question.
Whereas A) and B), are systematically available, C) is not.  
However, \cite{AHH-1}'s case of it is trivially solvable and trivially substitutable, so this is the method chosen in \cite{AHH-1} and the current Paper.  

\ni Procedure D) Some approaches -- covering at least a range of theories and of purposes -- make use of gauge-fixing prior to quantization.

\section{BSW action: use as counterexample and motivation in its own right}\label{BSW-Mot}

\ni {\bf Lemma} The adjectives `primary' and `secondary', as applied to constraints, are in fact formulation-dependent rather than theory-dependent.

\mbox{ } 

\ni I demonstrate that by a counterexample to formulation-independence of these adjectives.
Moreover, my choice for this counterexample is none other than the status of the GR Hamiltonian constraint itself.
In particular, I explain how this is primary in the Baierlein--Sharp--Wheeler (BSW) \cite{BSW} formulation of GR 
even though it is secondary in the more well-known Arnowitt--Deser--Misner (ADM) \cite{ADM} formulation of GR. 
The ADM formulation, following from the action\footnote{Here $\Sigma$ is a fixed spatial topology, 
taken to be compact without boundary in the current Paper since our more specialized model is on $\mathbb{S}^3$. 
$\mh_{ab}$ is the spatial 3-metric; these are the basic configurational variables in the geometrodynamical approach to GR, 
with corresponding redundant configuration space $\mbox{Riem}(\Sigma)$.
$\mh$, $D_i$ and $\mR$ are the corresponding determinant, covariant derivative and spatial Ricci scalar.
I include a cosmological constant $\Lambda$ since the closed homogeneous cosmology being perturbed about in the current Paper has need of this.  
$\upalpha$ is the lapse and $\upbeta^i$ is the shift.
$\mK_{ab}$ is the extrinsic curvature of the spatial slice $(\Sigma, \mh_{ab})$ within spacetime; its trace is denoted by $\mK$.  
In $\mS_{\tA\tD\tM}$, $\upbeta^i$ dependence, and further $\upalpha$ dependence, are hidden within $\mK_{ab}$ in accord with
$\mK_{ab} = \{\dot{\mh}_{ab} - 2\mD_{(a}\upbeta_{b)}\}/{2\upalpha}$. 
$\dot{\mbox{ }}$ is here $\pa/\pa \mt$, which in the case of GR is also a label time due to GR being already-parametrized.
$\mp^{ab}$ are the geometrodynamical momenta conjugate to $\mh_{ab}$, which are closely related to the extrinsic curvature: $\mp^{ab} = - \sqrt{\mh}\{ \mK^{ab} - \mK \mh^{ab} \}$.
Its trace is denoted by $\mp$.  \label{K}}

\ni \beq
\mS_{\sA\sD\sM} = \int \d\mt \int_{\Sigma}\d^3x\sqrt{\mh} \, \upalpha \{\mK_{ab}\mK^{ab} - \mK^2   +  \mR - 2\Lambda\} \mbox{ } ,
\label{S-ADM}  
\eeq
possesses both a scalar lapse coordinate $\upalpha$ and vector shift coordinates $\upbeta^i$.  
Both of these are Lagrange multiplier coordinates, and are indeed eventually among the multipliers used in appending constraints in the Dirac Algorithm at the Hamiltonian level.
As a consequence of being Lagrange multiplier coordinates, both have zero conjugate momenta: $\mp^{\upalpha} = 0$ and $\mp^{\upbeta}_i = 0$ as primary constraints. 
Then feeding these into the Dirac Algorithm, the GR Hamiltonian and momentum constraints -- respectively  

\ni\beq
{\cal H} := \mbox{$\frac{1}{\sqrt{\sh}}$}\big\{\mp^{ab}\mp_{ab} - \mbox{$\frac{\sp^2}{2}$}\big\} - \sqrt{\mh}\{\mR - 2\Lambda\} = 0 \mbox{ } ,
\eeq

\ni\beq
{\cal M}_i := -2\mD_j {\mp^j}_i = 0
\eeq
-- ensue as secondary constraints. 
In the BSW formulation -- following from the action

\ni\be
\mS_{\sB\sS\sW} = \int\d\lambda\int_{\Sigma}\sqrt{h}\sqrt{\mT\{\mR - 2\Lambda \}} \mbox{ } , \mbox{ } \mbox{ } 
\mT := \{\mh^{ac}\mh^{bd} - \mh^{ab}\mh^{cd}\}  \{\dot{\mh}_{ab} - 2\mD_{(a}\upbeta_{b)}\}  \{\dot{\mh}_{cd} - 2\mD_{(c}\upbeta_{d)}\} 
\label{S-BSW}
\ee
the statuses of $\upbeta^i$, $\mp^{\upbeta}_i = 0$ and ${\cal M}_i$ remain the same. 
However, the BSW formulation's action contains no lapse. 
Indeed, one way in which the BSW action arises is through multiplier elimination of the lapse: 
solving the corresponding Lagrange multiplier equation for the multiplier itself in order to eliminate said multiplier from the original (here ADM) action.
An interesting question then is whether and how the BSW formulation manages to encode ${\cal H}$, since in the ADM formulation this arose form the original presence of a lapse.

\mbox{ }

\ni The way the BSW action indeed still manages to encode ${\cal H}$ was envisaged in great generality by Dirac.
It occurs because the BSW action is homogeneous linear in the velocities.
So its momenta are homogeneous of degree 0 in the velocities.
Thus these are functions of ratios of velocities.
But there are one less independent such ratios than there are momenta.
Thus there must be at least one relation between the momenta themselves without any use made of the equations of motion.  
But by definition, this is a primary constraint.

\mbox{ }

\ni Then indeed, in the case of a `square root of a square' homogeneous linear action like BSW's, the corresponding primary constraint 
is in the form of a square of momenta equalling a momentum-less potential term \cite{RWR}. 
And in the particular case of the BSW formulation of GR, this is indeed a recovery of ${\cal H}$. 
Thus the BSW formulation succeeds in encoding the ${\cal H}$ which the canonical formulation of GR so requires. 
But clearly then in this case this constraint arises as a primary constraint, in contradistinction to how it arises as a secondary constraint in the ADM formulation. 
Hence by construction it is proven that primary or secondary are in fact formulation-dependent adjectives for constraints, rather than formulation-independent absolutes. $\Box$ 

\mbox{ }

\ni I give a degrees of freedom count for each of ADM and BSW in Appendix A.
The momentum constraint is formally interpreted as rendering the 3 local degrees of freedom of point-shuffling content of the 3-metric meaningless, 
the other 3 local degrees of freedom therein being termed the 3-geometry \cite{Battelle}. 
This amounts to passing from $\mbox{Riem}(\Sigma)$ to $\mbox{Superspace}(\Sigma) = \mbox{Riem}(\Sigma)/Diff(\Sigma)$ for $Diff(\Sigma)$ the 3-diffeomorphisms on $\Sigma$.  
In both the ADM and BSW cases, ${\cal H}$ and ${\cal M}_i$ close under the Poisson brackets to form the Dirac algebroid of constraints, by which the Dirac Algorithm successfully terminates. 
See e.g. \cite{Algebroid} for algebroids in general [these have structure function(al)s in place of Lie algebras' structure constants] 
and e.g. \cite{Dirac, T73, BojoBook} for the Dirac Algebroid in particular.

\mbox{ }

\ni Furthermore, BSW-type actions can be taken to arise from relational first principles `without ever passing though' the spacetime formulation of GR and the subsequent ADM split.  
In this way, relational first principles provide an answer to Wheeler's well-known question \cite{Battelle} 
which can be formulated as why the GR ${\cal H}$ takes the form it does without the above clause.  
In particular, the temporal relational principle implements the Leibnizian premise that `there is no time at the primary level for the universe as a whole'.
This is implemented by use of an action which firstly contains no extraneous time (such as Newton's) or time-like variable (such as ADM's lapse), 
and secondly is reparametrization invariant so as to not possess any kind of meaningful label time \cite{RWR, APoT3}. 
But `reparametrization invariant' is just another expression for homogeneous of degree one in the velocities! 
Hence the implementation is a BSW-type action, whence Dirac's argument above produces ${\cal H}$ as a primary constraint. 

\mbox{ }

\ni ${\cal H}$ can be regarded as an equation of time \cite{B94I} rather than an energy-type constraint, indeed being rearrangeable to form an emergent time expression at the secondary level 
(also interpretable as an emergent instant labeller or indeed emergent lapse).  
In this manner, Leibnizian primary timelessness is resolved by an implementation of Mach's insight that `time is to be abstracted from change' \cite{ACos2}, 
and an emergent kinematics in terms of a lapse as well as a shift arises, from which it is clear that the spacetime formulation of GR is recovered.
This is very interesting because it is a resolution of the classical precursor of the frozen formalism facet of the Problem of Time \cite{Kuchar92I93} in Quantum Gravity. 
Furthermore, this resolution is amenable to joint treatment of many of the other facets of the Problem of Time, as I have demonstrated in \cite{APoT2, APoT3, AM13, CapeTownTRiPoD}.  
(The present paragraph may only make sense upon realizing that almost all problem of time facets are already present at both the classical and semiclassical levels \cite{APoT2, APoT3}.)

\section{Motivation for this Paper's modelling assumptions}\label{Mot}

The preceding paragraph is my reason for interest in BSW-type actions, which is why I use them in \cite{AHH-1} and in then in the current Paper in further support of that.  
Having investigated Mechanics analogues of such \cite{FileR, AConfig}, 
I wanted to progress to model arenas exhibiting nontrivial GR spacetime relationalism, foliation issues and spacetime (re)construction. 
These are three further Problem of Time facets \cite{Kuchar92I93} which are not present in mechanics models and only trivially present due to homogeneity in minisuperspace models 
\cite{AMini}.  
In order to study these, I chose slightly inhomogeneous cosmology in a framework similar to Halliwell and Hawking's \cite{AHH-1}, based on Kucha\v{r}'s assertion \cite{Kuchar99} 
that these exhibited all the Problem of Time facets at the perturbative level. 
(Moreover, he never wrote out the evidence for this nor was working within a BSW-type action approach, so my workings are in any case new.)

\mbox{ }

\ni In doing this, I came across two unexpected phenomena 1) that the SVT split pieces of the constraints were no longer all behaving as first-class constraints. 
2) That the thin sandwich reduction -- i.e. addressing another Problem of Time facet as outlined in Section \ref{San} -- 
by itself did not send one down to superspace but rather to an intermediary space with extra dimensions per space point.  
I subsequently found that these two phenomena also occur in the vacuum counterpart of \cite{HH85}'s working, with the notable distinction of being simpler -- simple enough to resolve them, 
and as such I wrote this new Paper covering this simpler case for which those workings could be completed.
This could well be a useful guide in addressing the more cosmologically interesting case with scalar field included. 
It is furthermore useful as an example of two unexpected phenomena as well as a model which itself exhibits all of the Problem of Time facets, 
just not in as relevant a setting from the point of view of practical observational cosmology.

\mbox{ }

\ni Also note the difference between what my work and \cite{HH85} share - which do have scope for Background Independence and Problem of Time physics at the perturbative level, 
as per Kuchar's comment and my demonstration in \cite{AHH-1} -- and the more rigid QFT in curved spacetime based approaches of more widespread use in Cosmology.
The former have greater complexity than the latter (though there again, so do also supersymmetric and string approaches). 
In each case one reason to consider an extra level of complexity involves supplying perturbation spectra which then evolve in the later universe to good approximation within the latter scheme.
Semiclassical Quantum Cosmology then represents a moderate but highly universal choice of extra complexity within which to model such origin of structure, 
whilst retaining at least perturbative-sized realizations of the foundationally interesting Background Independence aspects and Problem of Time facets. 
That is a solid conservative approach which merits some attention, even though some others may choose to proceed differently as regards which layers of extra complexity to include 
in their models.
I next lay out this argument in the extended introduction, with a brief pointer after BSW actions have been introduced to their ties to relationalism, Background Independence 
and the Problem of Time.

\section{First-class status versus reformulations and manoeuvres}

A useful further piece of background is that whether `nontrivial consistent entities' are first-class constraints is also formulation-dependent.  
Certainly B) can be used to turn second-class constraints into first-class ones.
It is also clear that `manoeuvres' such as perturbation, symmetry or mode decomposition's pieces could turn second-class constraints into first-class ones by 
the obstruction to the closure of the classical brackets not having a contribution from the piece in question.  
First-class constraints producing other kinds of pieces, however, at first sight does not appear to be possible.
This is based on the thinking that if a constraint is not first-class it is second class, followed by the absurdity of the converse of the penultimate sentence.  
However, a way out of this is that what was a first-class constraint prior to a `manoeuvre' and is not after it need not now be a second-class constraint. 
This is because first-class and second-class is an exhaustive partition {\sl of constraints}, but the Dirac Algorithm permits one's equations to be entities other than constraints.
In other words, and staying within what is nontrivial and consistent, 
`manoeuvres' performed upon a first-class constraint are capable of producing pieces which are {\sl specifier equations}.

\mbox{ } 

\ni For the models in question (scalar field and vacuum) ${\cal H}$ and ${\cal M}_i$'s SVT pieces no longer all act as first-class constraints.
In retrospective, this is already clear in how ${\cal H}$ becomes {\sl three} conditions $^{\sS}{\cal H}$, $^{\sV}{\cal H}$ and $^{\sT}{\cal H}$; 
the manner in which ${\cal H}$ and ${\cal M}_i$ function as dynamical entities is changed by the SVT split.
If all of these pieces were first-class, they would then eat up more degrees of freedom than the original constraints did!   
The current Paper than attests to how in the simpler vacuum case with which I have made progress, this is indeed due to specifier equation production by the SVT split.
We shall see in the next Section how this is paralleled by the thin sandwich elimination being completed by taking out a piece of ${\cal H}$ as well as the naively expected entirety of 
${\cal M}_i$ so as to pass from the unreduced perturbative $\mbox{Riem}(\mathbb{S}^3)$ to the reduced perturbative $\mbox{Superspace}(\mathbb{S}^3)$. 
Note that the SVT pieces of the constraints ceasing to all be first-class constraints does not depend on the ADM versus BSW choice.

\section{The Thin Sandwich Problem as a particular type of reduction}\label{San}

Wheeler first contemplated a `thick sandwich': 
bounding bread-slice data to solve for the spacetime `filling' in between. 
This was in attempted analogy with Quantum Theory's Feynman path integral set-up for transition amplitudes between states at two different times \cite{WheelerGRT}, 
but failed to be well-posed.
Subsequently, Wheeler considered `thin sandwich' data to solve for a local coating of spacetime \cite{WheelerGRT}.
This is the `thin' limit of taking the bounding `slices of bread' -- the hypersurfaces $\mh_{ij}^{(1)}$ and $\mh_{ij}^{(2)}$ -- as knowns.
The equation to be solved is ${\cal M}_i$ in Lagrangian variables, and the unknown to be solved for is the shift vector $\upbeta^i$.
In more detail, the {\it thin sandwich PDE} is the BSW version of the above-described, i.e. containing what shall become an emergent lapse combination 
--the square root below -- rather than a primary lapse variable $\upalpha$ in this position: 

\ni\be
\mD_{j}\left\{ {\sqrt{\frac{2\Lambda - \mR}{\{\mh^{ac}\mh^{bd} - \mh^{ab}\mh^{cd}\}\{\pa{\mh}_{ab} - 2\mD_{(a}\upbeta_{b)}\}\{\pa{\mh}_{cd} - 2\mD_{(c}\upbeta_{d)}\}  }}}\right. 
\left.
\{\mh^{jk} \delta^{l}_{i} - \delta^{j}_{i}\mh^{kl}\}\{\pa{\mh}_{kl} - 2\mD_{(k}\upbeta_{l)}\}\right\} = 0 \mbox{ } . 
\label{Thin-San}
\ee
Note that in general this is not at all straightforward to solve or even to prove elementary things about, 
though Bartnik and Fodor showed unique solutions exist in a local sense \cite{TSC2}. 

\mbox{ } 

\ni Having done this, the emergent lapse can be computed (it clearly contains $\upbeta^i$ and thus cannot be computed priorly.
This can then be followed up by computing the extrinsic curvature of the incipient slice \cite{BSW}, by use of the formula in footnote \ref{K}.  

\mbox{ } 

\ni The emergent lapse can furthermore be reinterpreted in terms of an emergent time.
Despite \cite{BSW}'s title ``{\it Three-Dimensional Geometry as Carrier of Information about Time}", that paper does not itself cover this point.
However, Christodoulou did cover this in his ensuing paper ``{\it The Chronos Principle}" \cite{Christodoulou}. 
Moreover, that result lay unapplied and forgotten for many years until Barbour's rediscovery \cite{B94I} 
which subsequent papers have made built up on \cite{RWR, AM13} and indeed further interpreted as an implementation of Mach's Time Principle.  
Finally, from the expression for this classical Machian emergent time,\footnote{Here E denotes `extremization over the action functional', 
which is thus coincident with solving the Thin Sandwich Problem}
\ni\beq
t^{\se\sm}(x^a) = \mbox{\large E}_{\upbeta^i \mbox{ }  \in  \mbox{ }  \mbox{\scriptsize Diff}(\Sigma)        }
\int \left. \d \lambda
\sqrt{\{\mh^{ac}\mh^{bd} - \mh^{ab}\mh^{cd}\}  \{\dot{\mh}_{ab} - 2\mD_{(a}\upbeta_{b)}\}  \{\dot{\mh}_{cd} - 2\mD_{(c}\upbeta_{d)}\}} 
\right/   
\sqrt{     \mR - 2\Lambda    }  \mbox{ } ,
\label{tem}
\eeq
it is clear that approaches to the Problem of Time proceeding through the BSW action to get such an emergent time also require specifically the Thin Sandwich choice of PDE problem. 
This is ultimately why \cite{AHH-1} and the current Paper choose to approach the momentum constraint as a Thin Sandwich Problem.

\section{The Thin Sandwich Problem in slightly inhomogeneous cosmology}\label{San-SIC}

Let us next consider what becomes of the thin sandwich scheme in the vacuum slightly inhomogeneous cosmology model arena.
It has collapsed from a PDE system to algebraic equations (\ref{san-vac}): V and S pieces of ${\cal M}_i$, to be solved for the V and S pieces of the shift $\upbeta^i$ 
-- $j_{\sn}$ and $k_{\sn}$ respectively.\footnote{See Appendices C and D for the notation involved in this Section.} 
This is a well-determined system, and entirely straightforward to solve algebraically.  
Also note that -- in the collapse down to an algebraic equation -- the emergent versus primary lapse distinction is lost.  
This is since the lapse now occurs as an overall factor and thus can be cancelled out.
This permits greater parallels with an ADM treatment of the Lagrangian variables form of ${\cal M}_i$; \cite{HH85} did in fact provide such a parallel.
Differences are 1) they did not solve this for $j_{\sn}$, $k_{\sn}$, so theirs is not a thin sandwich scheme.
2) The perturbed part of the lapse $g_{\sn}$ enters their working 
(and an equation arises from varying with respect to that, so now one has a version based on a system of 4 equations in 4 unknowns). 
In contrast, my version is based upon the relational first principles approach to BSW actions not assuming lapse primality, and thus not having a $g_{\sn}$ in the first place, 
alongside requiring the thin sandwich's choice of variables to solve for.
This is not only due to Wheeler's arguments in posing the specific Thin Sandwich Problem, 
but also due to the thin sandwich choice of variables to be solved for then entering the expression (\ref{tem}) for $t^{\se\sm}$.
3) Finally, of course, \cite{HH85}'s version includes a scalar field, whereas the current Paper's is in vacuo.  
Moreover, \cite{AHH-1} considers the no primary lapse thin sandwich equation with scalar field, thus removing difference 3).  

\mbox{ } 

\ni Now, neither of these eliminations is accompanied by the line element losing a shift's amount of variables
so as to complete losing a $Diff(\mathbb{S}^3)$'s amount of variables.
Ie shift variables are being eliminated without taking partners away with them, which reflects that the equations being reduced cannot all be first class.
So, whereas (\ref{san-vac}) being algebraic parallels how relational particle models \cite{FileR, QuadI} deal with their analogue of the thin sandwich 
-- elimination of translation, rotation and optionally dilatation corrections -- however, there the constraints are all first-class, 
so each auxiliary eliminated is accompanied by the kinetic line element itself losing one dimension.
In a startling turn of events, this is {\sl not} emulated by the SIC thin sandwich equations, indicating that the pieces of ${\cal M}_i$ are not themselves first class 
despite coming from an object ${\cal M}_i$ which in its original unsplit form is widely known to be first-class.

\mbox{ } 

\ni Next note that the amount by which this fails differs between the vacuum and scalar field cases. 
In particular, unlike for the case including the scalar field, in the vacuum case the thin sandwich elimination additionally causes the scalar mode $a_{\sn}$ to drop out from 
the reduced kinetic term.
This is the first of two features by which the vacuum case is more straightforward to resolve than the case with scalar field.
To be specific, starting from the unreduced formulation's SVT split n-modewise (see Appendix C) configuration space line element (\ref{unred-vac}), 
the outcome of substituting in the solutions of the n-modewise thin sandwich equations (\ref{san-vac}) is the n-modewise reduced line element

\ni \beq
\mbox{$\frac{2}{\mbox{\scriptsize exp}(3\Omega)}$} \d s\mbox{}^2 = \{-1 + A_{\sn}\}\d\Omega^2 + \mbox{$\frac{2}{3}$}\d\Omega \, \d A_{\sn} + ||\d v_{\sn}||^2 \mbox{ } .  
\label{red-metric}
\eeq
This is a 5-$d$ line element, due to the $a_{\sn}$ coordinate having also dropped out, 
so this is now only short by 1 as regards removing a $Diff(\mathbb{S}^3)$ amount of degrees of freedom.
[5-$d$ for 3 degrees of freedom of n-modewise superspace, 1 coupled scale variable and the 1 that the calculation comes out short by.]
See Appendix D for the definitions of the $||v_{\sn}||$ and $A_{\sn}$ quantities that are useful in performing the reduction procedure.  
Finally note that $A_{\sn}$ is a mixed SVT quantity, in the sense of being a sum of S, V and T parts.

\mbox{ }

\ni Let us next analyze the geometry of this line element.
Computing the curvature tensors for (\ref{red-metric}) \cite{Maple} gives zeros all round. 
This establishing this metric to be an old friend in an unfamiliar guise -- 5-$d$ Minkowski spacetime -- and accorded also an unusual physical interpretation: 
as a reduced configuration space for the n-modewise slightly inhomogeneous cosmology on $\mathbb{S}^3$.  
This motivates finding a coordinate transformation which renders this into manifest 5-$d$ Minkowski spacetime form; this is

\ni \beq
t_{\sn} := \mbox{$\frac{2}{3}$}\sqrt{A_{\sn} - 1} \, \mbox{cosh}\big(\Omega + \mbox{$\frac{1}{3}$}\mbox{ln}(A_{\sn} - 1)\big) \mbox{ } ,
\eeq

\ni \beq 
w_{\sn} := \mbox{$\frac{2}{3}$}\sqrt{A_{\sn} - 1} \, \mbox{sinh}\big(\Omega + \mbox{$\frac{1}{3}$}\mbox{ln}(A_{\sn} - 1)\big) \mbox{ } \, \, .  
\eeq
Then indeed in these coordinates, 

\ni \beq
\mbox{$\frac{2}{\mbox{\scriptsize exp}(3\Omega)}$} \d s\mbox{}^2 = - \d t_{\sn}^2 + \d w_{\sn}^2 + ||\d v_{\sn}||^2 \mbox{ } .
\eeq
The cosh and sinh combination here just corresponds to the standard Rindler coordinates manoeuvre. 
The other individual transformations involved are just basic single-coordinate transformations and a diagonalizing transformation. 
Of course, the function sitting on the left-hand side signifies that the actual metric is but {\sl conformal} to Minkowski spacetime.
Also note that $||\d v_{\sn}||^2$ is the sum of an S piece and a T piece, but $t_{\sn}$ and $w_{\sn}$ are both mixed SVT since the naturally arising $A_{\sn}$ variable is. 

\mbox{ }

\ni $A_{\sn}$ (or $t_{\sn}$ and $w_{\sn}$) arising thus in the reduction illustrates that -- rather probably also contrary to popular beliefs -- 
{\sl reduction does not necessarily preserve the unreduced version of the theory's SVT split}.\footnote{Also reduction can use up one piece of such a split entirely; 
e.g. this is one manifestation of the V sector having no physical content at second order.
Nor does reduction in the case with scalar field \cite{AHH-1} respect the metric--matter split which the unreduced version manifests as a direct sum of the two sectors as exposited by 
Teitelboim \cite{Teitelboim}.}

\mbox{ } 

\ni In reducing, we also need to keep track of the corresponding n-modewise potential -- originally (\ref{W-n}) in the unreduced formulation. 
This also simplifies in the above variables, to 

\ni \beq
V_{\sn} = -\mbox{$\frac{\mbox{\scriptsize exp($\Omega$)}}{2}$}
\left\{
1 + s_{\sn}^2   -   \{\mn^2 - 1\} d_{\sn}^2   +   \mbox{$\frac{A_{\tn}}{3}$} 
\right\} 
+ \mbox{exp}(3\Omega)\Lambda\{1 + A_{\sn}\}  
\mbox{ } .
\label{V-red-1}
\eeq
\ni We are next faced with completing the reduction to n-modewise superspace.
Here we benefit from the second difference between the vacuum case and the scalar field case.
Namely that, in the vacuum case $^{\sV}{\cal H}$ becomes -- including pulling out an overall factor of $c_{\sn}^2$ by which this equation indeed also frees itself of its vector mode 
content, which is ultimately unphysical, whereby this is a helpful freeing -- 

\ni \beq
5\dot{\Phi}_{\sn}^{2} - 16 \dot{\Phi}_{\sn} \dot{\Omega} + 8\dot{\Omega}^{2} + \mbox{exp}(-2\Omega) = 0  \mbox{ }  
\eeq
for 

\ni\beq
\Phi_{\sn} := \mbox{exp}(3\Omega)\{1 + A_{\sn}\}/3 \mbox{ } .  
\eeq 
One can then take this as an equation for the thus only temporarily convenient mixed-SVT variable $A_{\sn} = A_{\sn}(\Omega)$.  
In this manner, in this particular example of reduction, the intrusion of an SVT-breaking variable is only transient. 
I.e. $A_{\sn}$ arises at the `halfway house' level which materializes due to the sandwich move not by itself attaining the expected amount of reduction to n-modewise superspace.
But then the subsequent reduction of the halfway house to the required n-modewise superspace end-product of the reduction eats up $A_{\sn}$.

The resultant 4-$d$ fully Diff($\mathbb{S}^3$)-reduced line element is then 

\ni \beq
\mbox{$\frac{2}{\mbox{\scriptsize exp}(3\Omega)}$} \d s\mbox{}^2 = \{-1 + f_{\sn}(\Omega)\}\d \Omega^2 + ||\d v_{\sn}||^2  \mbox{ } ,
\label{8}
\eeq
for 

\ni \beq
f_{\sn}(\Omega) := A_{\sn}(\Omega) + \mbox{$\frac{2}{3}\frac{\d A_{\tn}(\Omega)}{\d \Omega}$} \mbox{ } . 
\eeq 
(\ref{8}) is conformally flat; moreover, a new scale variable can be defined which absorbs the first term's prefactor: 
\ni\beq 
\zeta_{\sn} := \mbox{$\int$}\sqrt{f_{\sn}(\Omega) - 1}\,\d\Omega \mbox{ } ,
\eeq 
leaving, up to a conformal factor, the simplified line element

\ni \beq
\d s^2 = - \d\zeta_{\sn}^2 + ||\d v_{\sn}||^2 \mbox{ } .  
\label{superspace-ds}
\eeq
Note that the V-sector has been completely used up in the reduction, corresponding to its well-known lack of physical content.
$\pa/\pa \underline{v}_{\sn}$'s components $\pa/\pa s_{\sn}, \pa/\pa d^{\so}_{\sn}, \pa/\pa d^{\se}_{\sn}$ are then among the 10 conformal Killing vectors.\footnote{Following a point  
raised by one of the Referees, I clarify that this 10 is the 10 of Killing vectors of the 4-$d$ flat spacetime that this space is conformal to, 
as opposed to the also common occurrence of 10 as the maximal number of conformal Killing vectors in 3-$d$.}

\mbox{ }

\ni Finally the corresponding configuration space of the n-modewise inhomogeneities themselves is also clearly flat $\mathbb{R}^3$, 
with the $\underline{v}_{\sn}$ playing the role of Cartesian coordinates.

\section{Discussion of reduction and quantization schemes}

I next comment on the sense in which I use the word `reduction'.  
I mean reduction down to superspace.
In the full GR setting, this is usually taken to be the same as removing the linear constraints.
However, the preceding section illustrates how these two notions are capable of being distinct.  
Furthermore, treating non-linear constraints -- in particular the quadratic ${\cal H}$ -- distinctly from linear constraints has ramifications.
For instance, \K observables \cite{Kuchar93} are defined as those which commute specifically with first-class linear constraints.
Also, reduced quantization quite often means reducing out just the linear constraints at the classical level.\footnote{Some other physicists e.g. \cite{CP} use the term `reduced 
quantization' in cases which involve the Hamiltonian constraint as well, or indeed just the Hamiltonian constraint out of working on minisuperspace where there are no other constraints.
However I follow \cite{Kuchar92I93} in terming these other approaches `tempus ante quantum' instead.
This name follows since these solve the Hamiltonian constraint for a particular momentum variable: 
$\mp = \mH_{`\ttt\tr\tu\te'}(\mbox{the other variables})$ for $\mH_{`\ttt\tr\tu\te'}$ a candidate for the so-called `true' Hamiltonian.  
This isolated $\mp$ is then taken to be conjugate to an internal time $t^{\ti\tn\ttt}$ which is already present at the classical level, i.e. `ante quantum'.
This isolated $\mp$ subsequently produces an $i\hbar\pa/\pa t^{\ti\tn\ttt}$ at the quantum level (or its functional derivative equivalent, depending on the scope of the model.
Thus this approach has a time-dependent Schr\"{o}dinger type equation in place of a time-independent alias stationary or frozen one such as the Wheeler--DeWitt equation.}
Some works \cite{I76} talk of superspace quantization as well. 
Of course, in the habitually considered case in which the preceding sense of reduction and passage to superspace are interchangeable, the two notions coincide.
But since in the current Paper's model these two notions of reduction do not coincide, we need to call the scheme involved specifically superspace quantization.  
I also note that the difference between reducing out linear constraints and reducing down to superspace is not so large in the current Paper's case, 
since the quadratic equation also reduced out in the latter -- $^{\sV}{\cal H}$ indeed belongs to a V-sector which has no physical content.

\mbox{ }

\ni In any case, what is not being considered in the current Paper's approach is {\it Dirac quantization}, 
in the sense of the alternative scheme in which first-class linear constraints are promoted to form additional quantum equations and only then resolved.  
There are various reasons why I do not consider such a scheme, including that the current model has the good fortune of permitting classical-level reduction to superspace, 
and that keeping the SVT-split classical linear equation pieces is not strictly an option because I have demonstrated that these pieces are not all first-class constraints.
I gave more arguments for adopting reduced schemes when possible in \cite{APoT2, FileR} (based on further Problem of Time facets and operator-ordering considerations). 
It is furthermore well-known that reduced and Dirac quantization schemes often do not agree in detail; 
nor can we, for now at least, discern which of these schemes occurs in nature from semiclassical quantum cosmology imprints in observational cosmological data.
Due to this, physicists should be free to develop a variety of such schemes, including the current Paper's superspace quantization scheme.

\section{Outline of superspace quantization of slightly inhomogeneous cosmology}

The above flat metrics are useful in a number of applications.
As an example of this, starting afresh with the reduced configuration space line element (\ref{superspace-ds}) leads to a n-modewise superspace Hamiltonian

\ni \beq
\widetilde{\cal H} = \mbox{$\frac{1}{2}$}\{-p_{\sn}^{\xi \, 2} + ||\underline{p}^{v}_{\sn}||^2\} + \widetilde{V}(\xi_{\sn}, \underline{v}_{\sn}) \mbox{ } .  
\eeq
Quantizing this then gives 

\ni \beq
0 = \widehat{\widetilde{\cal H}} \, \Psi_{\sn} = \mbox{$\frac{\hbar^2}{2}$}\{\pa_{\xi_{\tn}}^2 - D^2\}\Psi_{\sn} + \widetilde{V}(\xi_{\sn}, \underline{v}_{\sn})\Psi_{\sn}  \mbox{ } . 
\eeq
This is mathematically in the form of a `spacetime position dependent mass' analogue of the Klein--Gordon equation 
(with the interpretation of the indefinite `spacetime' of course now being the slightly inhomogeneous n-modewise superspace configuration space).
Note that this equation still splits into scalar (i.e. now the scalar sum variable $s_{\sn}$: see Appendix C) and tensor parts. 
Each of these parts is coupled to the scale variable but not directly to each other.  

\mbox{ } 

\ni Moreover, as \K pointed out  \cite{Kuchar81Kuchar91}, a Klein--Gordon type equation having a non-constant mass term is in general a significant complication,  
due to this affecting the interpretation of the corresponding QM.  
Also note that equations of this form are familiar from e.g. Einstein--minimally coupled scalar isotropic, Bianchi I and diagonal Bianchi IX Quantum Cosmology \cite{Mini-QC}.  
Our point is that the regimes covered by such equations additionally extend to slightly inhomogeneous cosmology's split into n-modewise problems.  
Halliwell and Hawking's quantum scheme is of this nature too, though in that case of course accompanied by further quantum wave equations which are linear in their momenta.  
Instead of such, our superspace configuration space has coordinates whose physical meanings are somewhat more involved, but whose geometrical form is none the less itself simple. 

\mbox{ }

\ni There is a further commonness in the type of underlying mathematics involved. 
Both schemes are based on the same level of approximation which truncates terms at quadratic order, by which both schemes share the mathematics of multiple harmonic oscillators.  
As such, {\it leading-order} results as regards form of quantum solutions and structure formation are expected to be unaffected, 
while the differences between the two schemes are expected to show up in more detailed results.
Moreover, our scheme does not rely on hatting entities which have ceased to be playing the role of first-class constraints (for which how to Dirac-quantize is in fact rather less clear).
Thus there is a theoretical reason to consider our scheme's form of these more detailed results.

\mbox{ } 

\ni Paralleling the latter parts of \cite{HH85}, 
the scheme presented in this article straightforwardly admits a semiclassical regime as can be used to make quantum cosmological calculations and predictions. 
This shall be the subject of a subsequent Article \cite{AHH-3}.  

\mbox{ } 

\ni{\bf Acknowledgements} I thank to those close to me.  
I thank those who hosted me and paid for the visits: Jeremy Butterfield, John Barrow and the Foundational Questions Institute.
I thank also to Chris Isham and Julian Barbour for a number of useful discussions over the years, and also the anonymous referees.  

\begin{appendices}

\section{Counts of degrees of freedom}

\ni The ADM versus BSW counts of degrees of freedom are as follows. 
   10 $\times$ 2 degrees of freedom (based on 6 in $\mh_{ij}$, 3 in $\upbeta_i$ and 1 in $\upalpha$) 
--  3 $\times$ 2 (the $\mp^{\upbeta}_i$) 
--  1 $\times$ 2 ($\mp^{\upalpha}$)  
--  3 $\times$ 2 (the ${\cal M}_i$) 
--  1 $\times$ 2 (the ${\cal H}$) = 
    2 $\times$ 2 (based on 2 degrees of freedom per space point, matching the number of linearized-regime gravity wave polarizations).
Versus 
    9 $\times$ 2 degrees of freedom (based on 6 in $\mh_{ij}$, 3 in $\upbeta_i$ now with {\sl no} lapse) 
--  3 $\times$ 2 ($\mp^{\upbeta}_i$) 
--  3 $\times$ 2 (${\cal M}_i$) 
--  1 $\times$ 2 (${\cal H}$) = 
    2 $\times$ 2, so the overall outcome is in agreement with the previous.
This equivalence of outcome established, I stick to the current Paper's argued choice of BSW formulation, presenting the required wider range of cases for this in Fig 1. 

{            \begin{figure}[ht]
\centering
\includegraphics[width=0.65\textwidth]{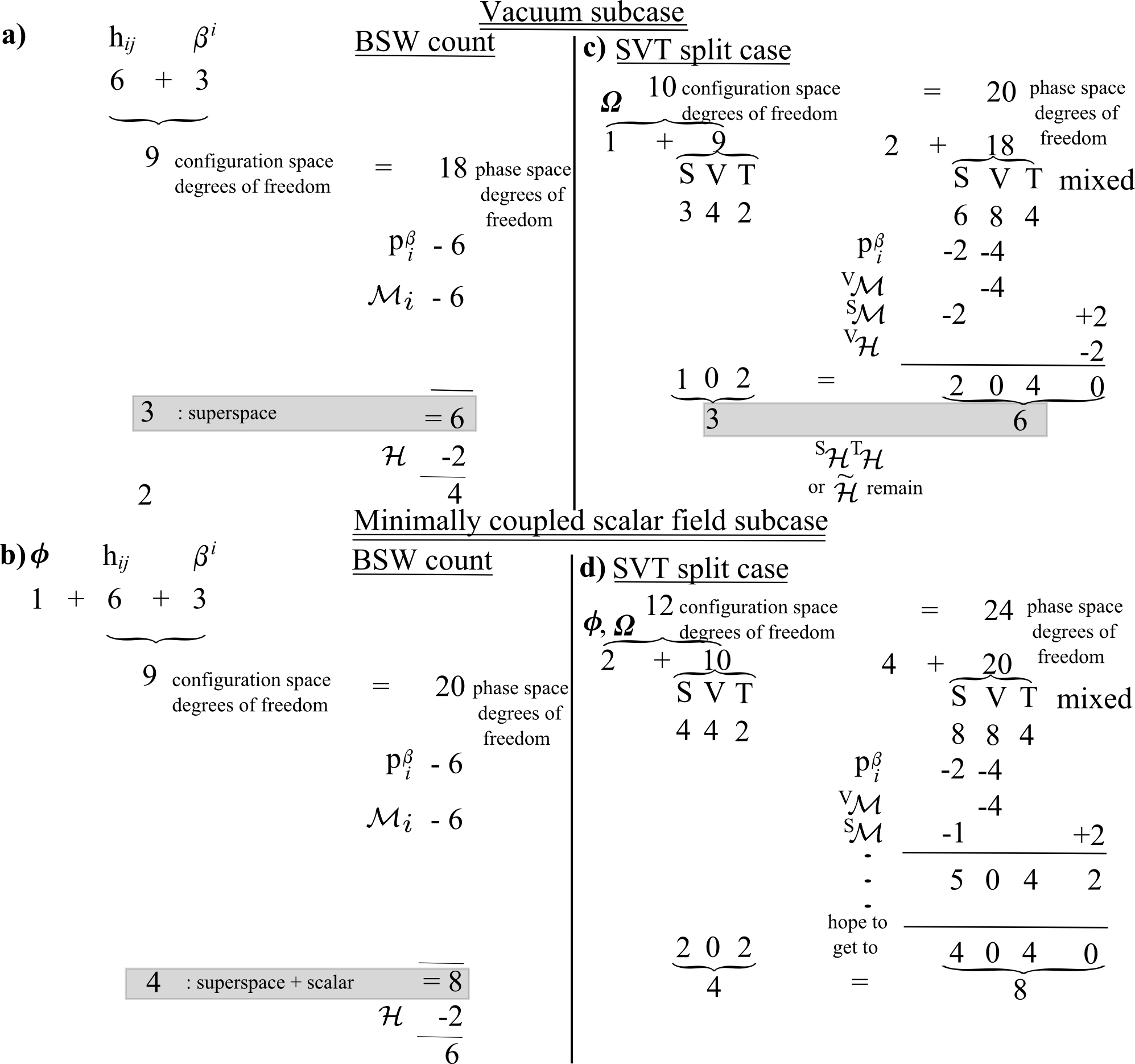}
\caption[Text der im Bilderverzeichnis auftaucht]{ \footnotesize{a) is the usual BSW analysis for vacuum GR.
b) is the version additionally including a minimally-coupled scalar field with homogeneous part denoted by $\phi$, 
as per \cite{AHH-1} and motivated by \cite{HH85} and standard cosmology making ample use of scalar fields.  
c) and d) then consider the outcome in the vacuum and scalar field case respectively, upon specializing to perturbatively inhomogeneous cosmology and applying the SVT split.
Note that each of c) and d) contain the Misner variable $\Omega$; this is because each n-modewise problem has this coupled to it.  
Also note that c) and d) exhibit the SVT split of the degrees of freedom in three columns, alongside marking a mixed SVT variable $A_{\sn}$ arising. 
In the vacuum case c), $A_{\sn}$ is itself subsequently reduced out, so the superspace presentation retains an ST split 
(the V sector, which is unphysical, has been fully used up by this stage).
Finally note how the momentum constraints fail to take out a $Diff(\mathbb{S}^3)$'s worth of degrees of freedom by an amount which is moreover different in each of the two cases.
The missing freedom has now been transferred to the multiple Hamiltonian constraints. 
I concentrate upon the descent to n-modewise Superspace($\mathbb{S}^3$); attaining this is marked with grey rectangles for clarity. 
This objective is presently attained for c) but not for d), hence the absense of any grey rectangle in d).  
This is due to the vacuum case c)'s $^{\tV}{\cal H}$ turning out to being able to eliminate the non SVT split variable $A_{\sn}$, 
which in this case therefore attains only a temporary significance in performing successive reductions.  
That the two cases c) and d) behave differently in this regard is an interesting result in its own right, due to the motivation in Secs \ref{BSW-Mot} and \ref{San}.} }
\label{SIC-Count} \end{figure}          }

\section{Caveats with BSW-type approaches}

A first caveat with BSW approaches is the fragility of the sandwich elimination move \cite{PR, Pitts}.
However, this does not apply in the desired context, since there the BSW-type relational action arises from relational first principles without there ever being a lapse to eliminate 
in the first place.

\mbox{ }

\ni A second caveat is that BSW-type actions might apply only locally \cite{TB, FileR}.  
This is one of the reasons why my approach which makes use of some such is to a local resolution of the Problem of Time. 
Local here means e.g. local in space, in configuration space and within spacetime (due to the previous two and the geometrodynamics to GR inter-relation).
This is due to there being some issues with BSW actions upon reaching zeros of its potential factor.\footnote{These are not severe issues like in the positive definite 
mechanics counterpart where these represent a halt to the motion, 
because with GR's supermetric being indefinite, they correspond to the motion going tangential to superspace's mathematical analogue of the light-cone. 
GR superspace's null cone, however, has no causality connotations, indeed Bianchi IX going through Kasner-like epochs is of this nature.  
Moreover, Bianchi IX going through such epochs is in the `approach to the Big Bang singularity' regime, whereas the current study 
is geared toward a later and more benign period -- the semiclassical quantum cosmological regime.
BSW-type actions remain suited to this regime, which still has just enough complexity to exhibit almost all of the Problem of Time phenomena (in the small perturbations).
Whether perturbative study of inhomogeneities within a BSW action will be trustworthy {\sl in all details} is then an interesting question. 
In this way, both caveats may eventually come to play some part, as may differences between reducing the ADM form of \cite{HH85} itself versus reducing the BSW form.  
However, the first few results that the current manuscript considers are not affected by these smaller differences, locality caveat and stability caveat.}
%
Solving the thin sandwich equation itself has locality caveats which parallel the above.

\section{SVT expansions and unreduced formulation of the paper's model}

The spatial 3-metric is here expanded as \cite{HH85}

\ni \beq
\mh_{ij} = \mbox{exp}(2\Omega(t))\{S_{ij}(t) + \upepsilon_{ij}(\underline{x}, t)\} \mbox{ } . \mbox{ } \mbox{ }  
\label{Sepsi}
\eeq
Here, $\Omega = \mbox{ln}\,a$ is the {\it Misner variable} for $a$ the cosmological scalefactor. 
$S_{ij}$ is the standard hyperspherical $\mathbb{S}^3$ metric.\footnote{I use straight font for spatially dependent quantities and italic font for spatially independent ones.}  
$\upepsilon_{ij}$ are inhomogeneous perturbations of the form \cite{F35GS78}

\ni $$
\upepsilon_{ij}=\sumnlm 
\big\{ 
\sqrt{\mbox{$\frac{2}{3}$}} a_{\sn\sll\sm}S_{ij}\mQ^{\sn}\mbox{}_{\sll\sm} + \sqrt{6}      b_{\sn\sll\sm}\{\mP_{ij}\}^{\sn}\mbox{}_{\sll\sm}         + 
\sqrt{2}\{c^{\so}_{\sn\sll\sm}\{\mS^{\so}_{ij}\}^{\sn}\mbox{}_{\sll\sm}    +         c^{\se}_{\sn\sll\sm}\{\mS^{\se}_{ij}\}^{\sn}\mbox{}_{\sll\sm}\} 
$$

\ni\beq
+  2 \{d^{\so}_{\sn\sll\sm}\{\mG^{\so}_{ij}\}^{\sn}\mbox{}_{\sll\sm}    +         d^{\se}_{\sn\sll\sm}\{\mG^{\se}_{ij}\}^{\sn}\mbox{}_{\sll\sm}\}  
\big\} .
\label{abcd}
\eeq
I subsequently use n indices as a shorthand for nlm.  
The     $\mQ_{\sn}(\underline{x})$                                                   are the                      $\mathbb{S}^3$             scalar harmonics, 
        $\mS_{i\,\sn}^{\so}(\underline{x})$  and $\mS^{\se}_{i,\sn}(\underline{x})$  are the transverse           $\mathbb{S}^3$             vector harmonics, 
and the $\mG_{ij\,\sn}^{\so}(\underline{x})$ and $\mG^{\se}_{ij,\sn}(\underline{x})$ are the transverse traceless $\mathbb{S}^3$ symmetric 2-tensor harmonics.
The superscripts `$\mo$' and `$\me$' for stand for `odd' and `even' respectively.  
The $\mS_{ij\,\sn}(\underline{x})$ are then given by $\mD_j\mS_{i\,\sn} + \mD_i\mS_{j\,\sn}$ (for each of the $\mo$, $\me$ superscripts, which I suppress post introduction). 
The $\mP_{ij\,\sn}(\underline{x})$ are traceless objects given by $\mP_{ij\,\sn} := \mD_j\mD_i\mQ_{\sn}/\{\mn^2 - 1\} + S_{ij}\mQ_{\sn}/3$. 
Additionally, the shift is expanded as

\ni\beq
\upbeta_i = \mbox{exp}(\Omega) \sum\mbox{}_{\mbox{}_{\mbox{\scriptsize $\sn, \sll, \sm$}}} 
\left\{
k_{\sn\sll\sm} \{\mP_i\}^{\sn}\mbox{}_{\sll\sm}/\sqrt{6} + 
\sqrt{2}  \{  j^{\so}_{\sn\sll\sm}  \{\mS^{\so}_i\}^{\sn}\mbox{}_{\sll\sm} + j^{\se}_{\sn\sll\sm}\{\mS^{\se}_i\}^{\sn}\mbox{}_{\sll\sm}  \}
\right\}
\eeq
for $\mP_{i\,\sn} := \mD_i\mQ_{\sn}/\{\mn^2 - 1\}$.  
The mode expansion coefficients $a_{\sn}$, $b_{\sn}$, $c_{\sn}$, $d_{\sn}$, and the shift expansion coefficients $j_{\sn}$, $k_{\sn}$, are all functions of $t$.
One can consider the above on an n by n basis, which I term an n-{\it modewise split}.

\mbox{ } 

\ni Then the vacuum case's n-modewise unreduced configuration space line element is 

\ni $$
\d s^2 =  \mbox{$\frac{   \mbox{\scriptsize exp}(3\Omega)  }{2}$} 
\left\{
-\d{a}_{\sn}^2 + \mbox{$\frac{\sn^2 - 4}{\sn^2 - 1}$}\d{b}_{\sn}^2 + \{\mn^2 - 4\}\d{c}_{\sn}^2 + \d{d}^2_{\sn}
+ \mbox{$\frac{2}{3}$}\d A_{\sn}\d{\Omega} + \{-1 + A_{\sn}\}\d{\Omega}^2  
\right\}
$$

\ni\beq
- \mbox{exp(2$\Omega$)}
\left\{
\{\mn^2 - 4\} \d{c}_{\sn} j_{\sn} + 
\left\{
\d{a}_{\sn} + \mbox{$\frac{\sn^2 - 4}{\sn^2 - 1}$}\d{b}_{\sn} 
\right\}
{k_{\sn}}/{\mbox{\scriptsize 3}}     
\right\} 
+ \mbox{$\frac{\mbox{\scriptsize exp}(\Omega)}{2}$}
\left\{
\{\mn^2 - 4\}j_{\sn}^2 - {k_{\sn}^2}/{\mbox{3\{n$^2$ -- 1\}}} 
\right\} \mbox{ } ,
\label{unred-vac}
\eeq
for $A_{\sn}$ given by (\ref{A-n}), and its n-modewise potential term is 

\ni \beq
V         = -\mbox{$\frac{\mbox{\scriptsize exp($\Omega$)}}{2}$}
\left\{
1 + 
\mbox{$\frac{1}{3}$}
\left\{
\mn^2 - \mbox{$\frac{5}{2}$}
\right\}
a_{\sn}^2 + \mbox{$\frac{\{\sn^2 - 7\}}{3}  \frac{\{\sn^2 - 4\}}{\sn^2 - 1}$}  b_{\sn}^2 + \mbox{$\frac{2}{3}$}\{\mn^2 - 4\}a_{\sn}b_{\sn} 
- 2\{\mn^2 - 4\}c_{\sn}^2 - \{\mn^2 + 1\}d_{\sn}^2  
\right\} 
+ \mbox{exp}(3\Omega) \Lambda\{1 + A_{\sn}\}  \mbox{ } .
\label{W-n}
\eeq
I have here included a cosmological constant term in addition to \cite{HH85}'s spatial Ricci scalar potential term.  
Finally, the thin sandwich equations (Lagrangian variables forms of $^{\sS}{\cal M}$ and $^{\sV}{\cal M}$) are  

\ni \be
\dot{a}_{\sn}  +  \mbox{$\frac{\sn^2 - 4}{\sn^2 - 1}$} \dot{b}_{\sn}  +  \mbox{$\frac{\mbox{\scriptsize exp($-\Omega$)}}{\mbox{\scriptsize n$^2$ -- 1}}$} k_{\sn}   = 0 
\mbox{ } , \mbox{ } \mbox{ }                        \dot{c}_{\sn}  -               \mbox{exp($-\Omega$)}                                                  j_{\sn}   = 0 
\mbox{ } .  
\label{san-vac}
\ee

\section{Some subsequently useful variables, with geometrical interpretation}

\ni Some variables that are useful in the reduced scheme are as follows.  
The {\it scalar sum variable} is

\ni\beq
s_{\sn} := a_{\sn} + b_{\sn} 
\eeq
up to a constant multiplicative factor.   
This is a \K observable (i.e. it commutes with the system's linear constraints ${\cal M}_i$).  
This combination also subsequently enters the definition of the Mukhanov--Sasaki variable \cite{MS} in the case with minimally-coupled scalar matter.

\mbox{ }

\ni I furthermore use $\underline{v}_{\sn}$ for the Cartesian 3-vector of inhomogeneities with components $[s_{\sn}, d^{\so}_{\sn}, d^{\se}_{\sn}]$.
$||\mbox{ }||$ is then the corresponding Euclidean norm on this n-modewise perturbatively inhomogeneous 3-$d$ space, and $D^2$ is the corresponding flat 3-$d$ Laplacian.

\mbox{ } 

\ni I also define 

\ni\beq
A_{\sn} := -\frac{3}{2}
\left\{
a_{\sn}^2 - 4
\left\{ 
\frac{\mn^2 - 4}{\mn^2 - 1}b_{\sn}^2 + \{\mn^2 - 4\}c_{\sn}^2 + d_{\sn}^2
\right\}
\right\} \mbox{ } .
\label{A-n}
\eeq
\ni This is the gravitational sector configuration space's {\it volume correction term}, 
in the sense of being the first perturbative correction to the expansion of the unreduced configuration space metric's determinant \cite{AHH-1}.
It is additionally the sole coupling to the minisuperspace degrees of freedom [c.f. the $\d\Omega^2$ term in (\ref{unred-vac})].  
This bears some resemblance to the 3-body problem's ellipticity variable ellip := $\rho^2_2 - \rho_1^2$, 
for $\rho_i$ the mass-weighted Jacobi relative inter-particle cluster separation vectors \cite{QuadI}.
Both are quadratic and comparisons by taking differences. 
On the other hand, ellip compares the base and median partial moments of inertia, 
whereas $A_{\sn}$ compares the amount of one of the scalar gravitational modes $a_{\sn}$ against that of the other gravitational modes.
All three of ellip, $s_{\sn}$ and $A_{\sn}$ drop out from reductions as `ubiquitous groupings', 
i.e. functional dependencies that widely feature throughout the corresponding theory's quantities.
Bigger such differences of quadratic expressions also drop out of the theory of larger planar $N$-body problems \cite{QuadI}.
Moreover, out of all these quantities $A_{\sn}$ alone subsequently vanishes from the working upon performing further reduction.
This goes hand in hand with $A_{\sn}$ alone not being a \K observable. 
This can be spotted since whereas the median and base being compared in ellip are themselves rotationally invariant entities, 
`the other gravitational modes' in $A_{\sn}$'s comparison include the unphysical vector modes $c_{\sn}$. 
Nor does the $A_{\sn}$ quantity's structure respect the status of $s_{\sn}$, rather than the individual $a_{\sn}$ and $b_{\sn}$, as an invariant quantity .

\mbox{ } 

\ni On the other hand, the final form of the inhomogeneous part of the vacuum potential (\ref{V-red-1}) is another quadratic difference,  
\beq
Q_{\sn} := s_{\sn}^2 - \{\mn^2 - 1\}d_{\sn}^2 \mbox{ } .
\eeq
This {\sl does} specifically compare the amount of the invariant scalar inhomogeneity $s_{\sn}$ with the amount of invariant tensor inhomogeneity $d_{\sn}$.
Moreover, in this case the individual pieces are already \K observables before the difference is taken. 

\mbox{ } 

\ni Finally, I note that $A_{\sn}$ already occurred within a suite \cite{Wada-Suite} 
of quadratic difference variables that were found to be useful variables for the quantum wavefunction to depend upon.  
This is both in its full form and in in various truncated forms, 
each of which amounts to comparing $a_{\sn}$ with various subsets of the other modes at the level of differences of sums of squares.
On the other hand, my study accounts for why these quantities occur in Wada's work by providing an underlying classical derivation of them, 
alongside a classical-level configuration space geometric meaning of $A_{\sn}$. 

\end{appendices}


\end{document}